\documentclass[graybox]{svmult}

\usepackage{mathptmx}       
\usepackage{helvet}         
\usepackage{courier}        
\usepackage{makeidx}         
\usepackage{graphicx}        
\usepackage{multicol}        

\makeindex             

\begin{document}
\title*{Fluids in Cosmology}

\author{Jorge L. Cervantes-Cota and Jaime Klapp}

\institute{J. L. Cervantes-Cota \and J. Klapp \at Departamento de F\'{\i}sica, Instituto Nacional de
Investigaciones Nucleares, ININ, Km 36.5, Carretera M\'exico-Toluca, La Marquesa 52750, Estado de M\'exico,
Mexico\\
\email{jorge.cervantes@inin.gob.mx , jaime.klapp@inin.gob.mx} \and
J. Klapp \at Departamento de Matem\'aticas, Cinvestav del Instituto
Polit\'ecnico Nacional (I.P.N.),
07360 M\'exico D. F., Mexico\\
\email{jaime.klapp@hotmail.com}}

%
%
\thispagestyle{empty} \maketitle
\setcounter{page}{69}

\abstract{
We review the role of fluids in cosmology by first introducing them in General Relativity and
then by applying them 
to a FRW Universe's model. We describe how relativistic and non-relativistic components evolve
in the background dynamics.
We also introduce scalar fields to show that they are able to yield an inflationary dynamics at
very early times (inflation) and
late times (quintessence). Then, we proceed to study the thermodynamical properties of the fluids
and, lastly,
its perturbed kinematics. We make emphasis in the constrictions of parameters by recent
cosmological probes.}

\section{Introduction \label{intro}}
Modern cosmology is understood as the study of fluids and geometry in the Universe. This task involves the development
of theoretical ideas about the nature of fluids and gravity theories, both to be compared with current observations that
cosmic probes have been undertaking.  The present understanding is condensed in the standard model of cosmology, that
incorporates the material content of the standard model of particle physics and Einstein's theory of General Relativity (GR)
with a cosmological constant. These two schemes, the fluid and gravity parts, have made predictions that have been tested
and confirmed, albeit there are still some issues that remain open. Certainly, we have really no firm knowledge of what
dark matter and dark energy are, as well as their nature
and detailed properties.  Still we are confident of some specific roles that these dark components play in
cosmology and astrophysics.  Their influence is at least gravitational, as so far we know from cosmic measurements.  This
knowledge alows us to build a picture of fluids in the background and perturbed geometry in the history of the
Universe and this is what we deal with in the present work.

The purpose of the present review is to provide the reader with
a panorama of the role that fluids play in the standard model of cosmology.
Tracking the recent history, in the late 1940's
George Gamow \cite{Ga46,Ga48} predicted that the Universe should had
begun from a very dense state, characterized by a huge density at
very high temperatures, a scenario dubbed the {\it Big Bang}, that was conjectured by George Lema\^{i}tre
in the early 30's. This scenario predicts that matter and light were at very high energetic
states in thermal equilibrium and described by a Planckian blackbody.
As the Universe expanded, it cooled down, and eventually matter and light decoupled.
The image of the last scattering of light is a fingerprint of the
initial state and remains today imprinted in the Cosmic Microwave Background Radiation (CMBR).
Gamow's scenario predicted that this primeval radiation would be measured at a temperature
of only a few Kelvin's degrees; since the expansion of the Universe cools down any density component.

The CMBR was for the first time measured by A. A. Penzias and R. W. Wilson in 1965 \cite{Penzias:1965wn}.
Later on, in the early 1990s
Smoot et al. \cite{Sm92} and Mather et al. \cite{Mather90} measured further important properties of this
radiation: its tiny anisotropies
for large angular scales and its blackbody nature. The first property -- also imprinted in the matter distribution --
accounts for the perturbed
fluids in the Universe that led to structure formation in the cosmos. The second property is a distinctive
sign of the equilibrium
thermodynamic properties of the primeval plasma -- composed of photons, electrons, and baryons, plus
decoupled (but gravitationally coupled) neutrinos, dark matter, and dark energy. The evolution and effects of
these fluids is the main concern of the present review.

We begin our work by explaining the context of fluids in GR, and especially in cosmology. We then analyze
the evolution of
perfect fluids - since real fluids allow them to be described as such- and their background dynamics.
We explain that the
main cosmic components are baryons, photons, neutrinos, dark matter, and dark energy.  We also introduce scalar
fields since
they are ubiquitous in modern cosmology because they enable to model different cosmic dynamics, from
inflation \cite{Gu81,Li90} and
dark energy \cite{Caldwell:1997ii,CoSaTs06} to dark matter \cite{Matos:1999et,Magana:2012ph}. Then, we proceed to
study the  thermodynamical properties of the fluids (as in Ref. \cite{CervantesCota:1900zz}) and, lastly, its
perturbed kinematics. We make
emphasis on the constraints of parameters as imposed by recent cosmological probes.

In this work, we use ``natural'' units $\hbar = c = k_{B} = 1$ and our geometrical sign conventions are as
in Ref. \cite{MiThWh73}.

\section{Fluids in general relativity \label{fluids_GR}}

The GR theory is based on the Einstein-Hilbert Lagrangian density
\begin{equation}
{\cal L} = \frac{1}{16 \pi G} (R + L_{m}) \sqrt{-g} \,\ ,
\label{elsc}  \end{equation}
where $R$  is the Ricci scalar, $G$ the Newton constant,  $g=|g_{\mu \nu}|$ is
the determinant of the metric tensor, and $L_{m}$ is the material Lagrangian that will give rise to the fluids.
By performing the metric variation to this equation, one obtains the
well known Einstein's field equations
\begin{equation}
 {\bf R}_{\mu \nu} -\frac{1}{2}R \,  {\bf g}_{\mu \nu} =  8\pi G  {\bf T}_{\mu \nu} \,\ ,
 \label{eesc}   \end{equation}
where $ {\bf R}_{\mu \nu}$ is the Ricci tensor and
$ {\bf T}_{\mu \nu}$ is the stress
energy--momentum tensor whose components are given through
$ T_{\mu \nu} \equiv - \frac{2}{\sqrt{-g}} \frac{\partial L_{m} \sqrt{-g}}{\partial g^{\mu \nu}}$.
Tensors in Eq. (\ref{eesc}) are symmetric which is a requirement of the
theory.  Being space-time four dimensional the imposed symmetry implies that
Eq. (\ref{eesc}) represents a collection of ten coupled partial differential equations. However, the
theory is diffeomorphism invariant, and one adds to them a
gauge condition, implying in general four extra equations to Eq. (\ref{eesc}) that reduce the
physical degrees of freedom. Thus, symmetries and gauge choice determine the fluid properties
allowed by the theory.

The stress energy-momentum tensor ${\bf T}$ encodes the information of the fluid, and all kinds of energy types
contribute to curve space-time: density, pressure, viscosity, heat, and other physical quantities. But before
introducing them, one
needs other elementary concepts.

Giving some reference frame, one defines the four-velocity
$\vec{u} \equiv d\vec{x}/d\tau$ as the vector tangent to the
worldline of a particle, with $\vec{x}$ being the local coordinates and $\tau$ the proper time along the worldline;
its four-momentum
is $\vec{p} = m \vec{u}$, where $m$ is the rest mass of the particle.
Now, given a space-time surface, $x^{\alpha}=$const., one defines its associated
one-form as $\tilde{dx^{\alpha}}$, to obtain the components
${\bf T} (\tilde{dx^{\alpha}}, \tilde{dx^{\beta}}) =  T^{\alpha \beta}$, which is interpreted as the flux
of momentum $\alpha$,
$p^{\alpha} = < \tilde{dx^{\alpha}}, \vec{p} >$, passing through the surface $x^{\beta}=$const. In this way,
$T^{00}$ is the energy density, which is the flux of momentum ($p^{0} =$particle's energy) that crosses the
surface $x^{0} =t=$const. and $T^{0 i}$ is the flux of energy that crosses the
surface $x^{i} =$const.; where latin labels run from 1 to 3 and greek labels from 0 to 3. Given the symmetry
of the tensor, $T^{i 0} =T^{0 i}$, that is, energy fluxes are equal to momentum densities since
mass equals relativistic energy. Finally, the components $T^{i j}$ denote
the momentum flux $i$ crossing the surface $x^{j}=$const., and again symmetry implies that $T^{ij}=T^{ji}$,
avoiding a net intrinsic angular momentum.

The left-hand side of Eq. (\ref{eesc}) is known as the Einstein tensor (${\bf G}_{\mu \nu}$) and, giving
the symmetries
of the theory, it happens to fulfill the Bianchi identities, that is, its covariant derivative is null.
This in turn implies, on the right-hand side,
a conservation law for any fluid within this theory. The conservation law reads:
\begin{equation} \label{cons-law1}
{\bf T}_{\mu \,\ ;\nu }^{~ \nu } = 0 .
\end{equation}
As we shall see, this equation is the most important since it encodes the thermodynamic laws of matter.

\section{Fluids in cosmology \label{sc}}

The kinematical properties of a fluid element are determined by its velocity, acceleration, shear, and
vorticity.   All these quantities are defined in the space-time, and for convenience one uses
comoving coordinates, that is Lagrangian coordinates that follow the flow motion. We refer the reader
to standard gravity textbooks for details, e.g., Refs. \cite{MiThWh73,Schutz:1985jx}. One splits the space-time
structure into surfaces of simultaneity to rest frame observers, with a projected metric on the surface
${\rm h}_{\mu \nu} = g_{\mu \nu} + u_{\mu} u_{\nu}$; where $u^{\mu}$ are the components of the four velocity
$\vec{u}$. In this frame it is natural to define
an expansion tensor, $\Theta_{\mu \nu} = \Theta_{(\mu \nu)} = \nabla_{(\mu} u_{\nu)}$, and
the vorticity tensor, $\omega _{\mu \nu} = \omega_{(\mu \nu)} = \nabla_{[\mu} u_{\nu]}$, where $\nabla$
operates on the projected 3-dimensional space.    The trace of the expansion tensor is a scalar measure
of the volume expansion, given by $\Theta=  \nabla_{\mu} u^{\nu}$, and the shear tensor is the projected symmetric
free-trace part of $\Theta_{\mu \nu} $, such that
$\Theta_{\mu \nu} = \sigma_{\mu \nu}  + \frac{1}{3} \Theta \: {\rm h}_{\mu \nu}$ (see Ref. \cite{ellis2012}).

Accordingly, the energy-momentum tensor associated to the fluid can be separated into components parallel
and orthogonal to the four velocity as:
\begin{equation} \label{gen-emt}
T_{\mu \nu} = \rho  u_{\mu}   u_{\nu} + q_{\mu}   u_{\nu} + q_{\nu}   u_{\mu} + P \, {\rm h}_{\mu \nu} + \pi_{\mu \nu} ,
\end{equation}
where $\rho = T_{\alpha \beta}  u^{\alpha} u^{\beta}$ is the energy density that includes rest masses and
possibly the internal
energy, such as the chemical energy; $P = {\rm h}^{\alpha \beta} T_{\alpha \beta}/3$
is the pressure; $q_{\mu}  = - {\rm h}^{\alpha}_{\mu} T_{\alpha \nu} u^{\nu}$ is the momentum density or energy flux
due to either diffusion or heat conduction; and
$\pi_{\mu \nu} = [ {\rm h}_{(\mu}^{\, \, \, \alpha} {\rm h}_{\nu)}^{\, \, \, \beta} -
\frac{1}{3} {\rm h}_{\mu \nu} {\rm h}^{\alpha \beta} ]T_{\alpha \beta}$ is the trace-free
anisotropic stress tensor due to viscosity.

A perfect fluid is an inviscid fluid with no heat conduction, that is, $q_{\mu} =0$ and $\pi_{\mu \nu} =0$. It is
analogous to an ideal gas in standard thermodynamics. In terms of the full metric, it is a standard practice to
represent it as:
\begin{equation} \label{pf-emt}
T^{\mu \nu} = (\rho + P)  u^{\mu}   u^{\nu} + P g^{\mu \nu} ,
\end{equation}
in comoving coordinates, $u^{\mu} = \delta^{\mu}_{0}$. Equation (\ref{pf-emt}), is the energy-momentum
tensor that correctly describes fluids in the background geometry of the Universe.

\section{Fluids in the standard model of cosmology \label{sc}}

The Universe is described by its material components and geometry. The former is fed with microscopic or
thermodynamic information
about the fluids and the latter is determined by Eq. (\ref{eesc}). In the following, we explain the features of
the geometry and the properties of the fluids that have governed the evolution of the standard model of
cosmology.

The {\it cosmological principle} states that the Universe is both spatially homogeneous and isotropic on large scales,
and this imposes a symmetry on the possible fluids present in it. Any departure from this symmetry in the fluid would be
reflected in the geometry through Eq. (\ref{eesc}). The symmetry assertion is compatible with observations
made of the all-sky cosmic microwave background radiation from the last twenty years, through the satellites COBE
\cite{Sm92} in the
1990s, the Wilkinson Microwave Anisotropy Probe (WMAP) \cite{Bennett:2012fp,Hinshaw:2012aka}
in the 2000's, and the PLANCK \cite{Ade:2013ktc} nowadays, although
some large scale CMBR anomalies in the isotropy have been detected \cite{Ade:2013nlj} that require further
investigation.  On the other hand, homogeneity and isotropy have also been tested for the distribution of matter
at large scales, see for instance Refs. \cite{Hoyle:2012pb,Marinoni:2012ba}.

In GR, as in any other metric theory, symmetries of the physical system are introduced through the metric tensor.
The homogeneous and isotropic space-time symmetry was originally studied by Friedmann, Robertson, and
Walker (FRW) (see Refs. \cite{Fr22,Fr24,Ro35,Ro36,Ro36a,Wa37}). The symmetry is encoded in and defines the unique form
of the line element:
\begin{equation}
 ds^{2} =g_{\mu \nu} dx^{\mu} dx^{\nu} = -dt^{2} + a^{2}(t) \,\  \left[ \frac{dr^{2}}{1-k r^{2}}
 + r^{2} (d\theta^{2} + {\rm sin}^{2} \theta \, d\phi^{2}) \right] ,
 \label{frwmsc} \end{equation}
where $t$ is the cosmic time, $r$, $\theta$, and $\phi$ are polar coordinates, and the constant
curvature can be adjusted to take the values $k$ = 0, +1, or $-1$ for a flat, closed, or open space,
respectively. $a(t)$ is the unknown
potential of the metric that encodes the size at large scales, and more formally, it is the {\it scale factor}
of the Universe that
measures how the model grows or shrinks as time evolves. Measurements show that it always grows, but a bounce in the
very early or final stages is possible (see, e.g., Ref. \cite{DeSantiago:2012nk}).

The beautiful symmetric FRW solutions to the Einstein Eqs. (\ref{eesc}) represent a cornerstone in
the development of modern cosmology, since with them it is possible to
understand the expansion of the Universe. Although in the first years of relativity, Einstein sought for a static
solution -- since
observations seemed to imply that -- it was soon realized by E. Hubble and others in the mid 1920's that the
Universe is indeed expanding, following Hubble's law \cite{Hu27}.

Using the FRW metric and a perfect fluid, the GR cosmological field equations are,
\begin{equation}
H^{2} \equiv \left(\frac{\dot{a}}{a}\right)^{\! 2}= \frac{8 \pi G}{3} \rho -
\frac{k}{a^{2}}
 \label{frw1sc} \end{equation}
and
\begin{equation}
\frac{\ddot{a}}{a} = - \frac{4 \pi G}{3} (\rho + 3 P) \,\ ,
\label{frw2sc} \end{equation}
where $H$ is the {\it Hubble parameter} that has dimensions of inverse of time, and therefore, it encodes the
model's expansion rate ; $H^{-1}$ is proportional to the age of the Universe. Moreover, $\rho $ and $P$ are the
density and pressure that enter in Eq. (\ref{pf-emt}). Dots
stand for cosmic time derivatives.

As explained above, the energy-momentum tensor is covariantly conserved, as shown by Eq. (\ref{cons-law1}). In the
present case, it implies the continuity equation,
\begin{equation}
\dot{\rho} + 3 H (\rho + P) = 0 \,\ .
 \label{frw3asc}
\end{equation}

Equations (\ref{frw1sc}), (\ref{frw2sc}), and (\ref{frw3asc}) involve three unknown
variables ($a$, $\rho$, $p$) for three equations, but the system is not mathematically closed, since the equations
are not all linearly independent, but just only two of them. Thus, an extra assumption has to be made to solve
the system.  The answer  comes from the micro-physics of the fluids considered. For the moment let us assume
a barotropic equation of state that is characteristic for different cosmic fluids, i.e.,
$w = {\rm const.}$ so that
\begin{equation}
\frac{P}{\rho} = w =
\left\{ \begin{array}{l@{\quad {\rm for} \quad}l}
\frac{1}{3}     &    {\rm radiation ~or ~relativistic ~matter,}\\
   0            &    {\rm dust,}\\
   1            &    {\rm stiff ~fluid,}\\
 - 1            &    {\rm cosmological ~constant ~or ~vacuum ~energy,}  \end{array} \right.
\label{ecsc} \end{equation}
to integrate Eq. (\ref{frw3asc}), yielding
\begin{equation}
\rho = \frac{M_{w}}{a^{3(1+w )}}  ~~~~~~~{\rm or} ~~~~~~
\frac{\rho_i}{\rho_{i0}} = \left( \frac{a_0}{a} \right)^{3(1+w_{i} )}  \,\ ,
 \label{frw3bsc}
\end{equation}
where $M_{w}$ is the integration constant and has different dimensions for different
$w$-fluids. The equation on the right shows a different re-scaling of the integration constant,
where the subscript $i$ stands for the different $i$-fluids.
Quantities with either a subiscript or superscript ``0'' are evaluated at the present time.
With this equation the system is mathematically closed and can be solved.

The system of ordinary differential equations described above needs a set of
initial, or alternatively boundary conditions to be integrated. One has to
choice a set of two initial values, say,
$(\rho(t_{*}), \dot{a}(t_{*}) ) \equiv (\rho_{*}, \dot{a}_{*})$
at some (initial) time $t_{*}$, in order to determine its evolution. A full analysis of this
assumption can be found in many textbooks \cite{MiThWh73,We72,We08}. In order to show some
some physical consequences of the early Universe, we assume that $k=0$. This is consistent with
data from recent cosmological probes, as we shall explain shortly. This tell us that curvature has not
played a role for most of the age of the Universe. On the other hand, this can be justified
as follows: from Eqs. (\ref{frw1sc}) and (\ref{frw3bsc}) we may see
that the  expansion rate, given by the Hubble parameter, is
dominated by the density term as $a(t) \rightarrow 0$, since
$\rho \sim 1/a^{3(1+w )} > k / a^{2}$ for $w > -1/3$, that is,
the flat solution fits very well the very beginning of times.
Therefore, taking $k =0$, Eq. (\ref{frw1sc}) implies that
\begin{eqnarray}
a(t) & = & [6 \pi G M_{w} (1+w)^{2}]^{\frac{1}{3(1+w )}}
(t-t_{*})^{\frac{2}{3(1+w )}} \nonumber \\[5pt]
 & = & \left\{ \begin{array}{l@{\quad {\rm for} \quad}l}
 (\frac{32}{3} \pi G M_{\frac{1}{3}})^{1/4}~(t-t_{*})^{1/2} & w=\frac{1}{3} ~
{\rm radiation,}\\
 (6 \pi G M_{0} )^{1/3} ~~~ (t-t_{*})^{2/3}    & w=0 ~ {\rm dust,}\\
 (24 \pi G M_{1} )^{1/6} ~~ (t-t_{*})^{1/3}     & w= 1 ~{\rm stiff ~fluid,}\\
\end{array} \right. \label{frwsolsc}\end{eqnarray}
and
\begin{equation}
a(t) =  a_{*} e^{H t}  \qquad {\rm for}\qquad w=-1 ~{\rm cosmological ~constant } ,
\label{isc} \end{equation}
where quantities with subscript ``$*$'' are integration constants,
representing quantities evaluated at the beginning of times, $t=t_{*}$. It is thought that within a
classical theory (as GR)
this initial time is at most as small as the
Planckian time ($t_{Pl} =10^{-43} $s), since prior to it GR has to be modified to include quantum effects. To
obtain Eq. (\ref{isc}), the argument given above to neglect $k$
is not anymore valid, since here $\rho= {\rm const.}$; that is,
from the very beginning it must be warranted that
$H^{2} \approx \frac{8 \pi G}{3} \rho_{*} > k/a^{2}_{*}$, otherwise
$k$ cannot be ignored. Nevertheless if $\Lambda$ is present, it will eventually dominate
over the other decaying components, this is the so-called {\it cosmological no-hair theorem} \cite{ChMo94}.
A general feature of all the above solutions is that they are expanding, at different Hubble rates,
$H= \frac{2}{3(1+w)} \frac{1}{t}$ for Eqs. (\ref {frwsolsc}) and $H = $const. for Eq. (\ref {isc}).

From Eq. (\ref{frwsolsc}) one can immediately see that at
$t=t_{*} , \,\ a_{*} = 0$ and from Eq. (\ref{frw3bsc}),
$\rho_{*} = \infty$, that is,
the solution has a singularity at that time, at the beginning of the Universe.
This initial cosmological singularity is precisely the Big Bang singularity.  As
the Universe evolves the Hubble parameter goes as $H \sim 1/t$, i.e.,
the expansion rate decreases, whereas the matter-energy content acts as an
expanding agent [cf. Eq. (\ref{frw1sc})]. It decelerates the expansion,
however, by decreasing asymptotically [cf. Eqs. (\ref{frw2sc}) and
(\ref{frw3bsc})]. In this way,
$H^{-1}$ represents an upper limit to the longevity of the Universe; for instance,
$H^{-1}=2t$ for $w =1/3$ and $H^{-1}=3t/2$ for $w =0$, $t$ being the
age of the Universe.

The exponential expansion (\ref{isc}) possesses no singularity (at finite times), being  the
Hubble parameter a constant.  A fundamental ingredient of this inflationary solution
is that the right-hand side of  Eq. (\ref{frw2sc}) is positive,
$\ddot{a}>0$, and this occurs when $\rho+3 p < 0$, that is, one
does not have necessarily to impose the stronger condition $w=-1$, but it suffices that $w <-1/3$,
in order to have a
moderate inflationary solution; for example,
$w =-2/3$ implies $a=a_{*} t^{2}$: a mild power-law inflation.

\begin{figure}[ht]
\includegraphics[width=10cm]{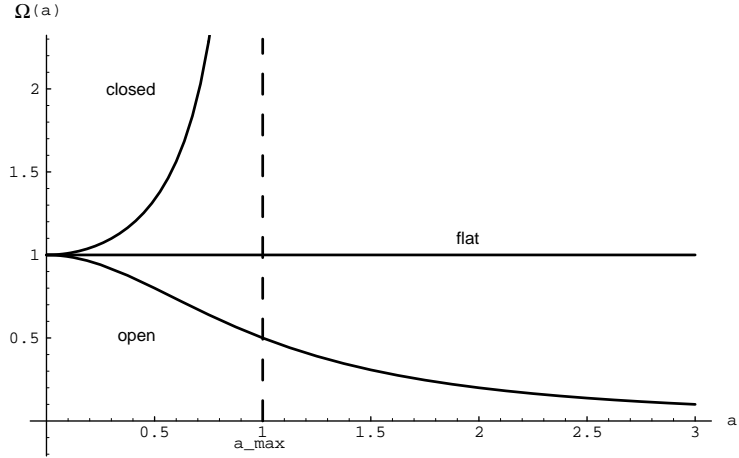}
\caption[The density parameter as a function of the scale factor]{The
parameter $\Omega$ as a function of the scale factor, $a$, in
a radiation dominated Universe (the dust model behaves similarly). For closed
models, with $k=+1$, $\Omega$ diverges as the
scale factor tends to its maximum value, whereas for open
models, with $k=-1$, $\Omega$ tends asymptotically to zero as the Universe
expands. Finally, for a flat metric, with $k=0$, $\Omega$ always remains
equal to one.
\label{omegasc} }
\end{figure}

It is convenient to define dimensionaless density parameters as $\Omega_i \equiv \frac{8 \pi G \rho_i}{3 H^2}$.
With them,
Eq. (\ref{frw1sc}) can be expressed as the constraint:
\begin{equation} \label{Omega}
\Omega \equiv \Omega_R +  \Omega_M + \Omega_\Lambda  = 1 + \frac{k}{a^{2}
H^{2}} ,
\end{equation}
where $i$ labels the different components present in the Universe: $R$ stands for the radiation components
(photons, neutrinos, and
relativistic particles), $M$ for matter which is composed of dark matter (DM) and baryons, and
$\Lambda$ for a cosmological constant. The actual values of the density parameters
$(\Omega_{R}, \Omega_{m}, \Omega_{\Lambda})$ impose a value for
the curvature term. If $\Omega > 1$, it turns out that $k$ is greater
than zero, meaning a Universe with a positive, closed curvature. If $\Omega < 1$, then $k < 0$, which corresponds
to a negative, open curvature. Obviously, a critical value is obtained
when $\Omega = 1$, then the spatial curvature is null, $k=
0$. The value of the energy density for which
$\Omega = [\rho + \Lambda/(8 \pi G)]/\rho_{c}  = 1$ holds is known as the {\it critical
density}, $\rho_c \equiv 3 H^{2}/8 \pi G$. The last term in  Eq. (\ref{Omega}) can be defined
as $ \Omega_{k} = -k/(a^{2} H^{2})$, and thus the Friedmann equation becomes a
constraint for the density parameters, i.e., $\sum_i \Omega_i =1$, and this expression holds at
any time. It is worth mentioning that solutions $\Omega(a)$ are unstable in the presence of a
curvature term (see Fig. \ref{omegasc}). In fact,
this is related to the flatness problem in the old cosmological picture: Why the Universe is nowadays
close to a flat model? Inflation offered the solution to this issue.

In (background) cosmology, typical times and distances are determined mainly by the Hubble parameter,
and in practice measurements
are often related to {\it redshift}, as measured from stars, gas, etc. It is then useful to express the
individual density parameters in terms of
the redshift ($z$), $1+z \equiv a_0/a(t)$, where $a_{0}$ is the scale factor at present and
is set to unity by convention. Today $z_0 =0$ and towards the early Universe the redshift grows.
In terms of the redshift the density parameters are, from  Eq. (\ref{frw3bsc}),
\begin{equation} \label{om-z}
\Omega_i = \Omega_{i}^{(0)} (1+z)^{3(1+w_i)} \; ,
\end{equation}
where $w_i$ is the equation of state parameter for each of the fluids considered. Now, the Hubble parameter
can be put in terms of the density
parameters. In the standard model of cosmology, considering
baryons, photons, neutrinos, cold dark matter (CDM), and a cosmological constant ($\Lambda$) -- termed
$\Lambda$CDM --, one has:
\begin{equation} \label{hz1}
H^2 = H_{0}^2  \sum_i \Omega_{i}^{(0)} (1+z)^{3(1+w_i)} \; .
\end{equation}

As defined above, the density parameter depends on
$1/H^2$, so to avoid a bias with the expansion rate one defines the {\it physical}
density parameter $\omega_i \equiv \Omega_i h^2$, where $h$ is a dimensionless number given by the Hubble constant
$H_{0} \equiv 100h$ km s$^{-1}$ Mpc$^{-1}$. The physical density parameters of matter are important since they are
directly determined from CMBR experiments. The current best-fit values for the physical density parameters
from PLANCK are \cite{Ade:2013zuv}:
$\omega_{b} = 0.022$, $\omega_{DM} = 0.120$, from which one computes the best fits
\begin{equation} \label{planck_params}
\Omega_b^{(0)} =  0.0492,  \quad \Omega_{DM}^{(0)} =  0.267,   \quad  \Omega_{\Lambda}^{(0)} =  0.683,  \quad h=0.671.
\end{equation}
The Universe at present is dominated by dark energy, which accounts for 68$\%$ of the energy budget, dark matter for
27$\%$, and in minor proportion baryonic matter only for about 5$\%$, from which visible matter is made of. Photons
and neutrinos
contribute in a much less proportion at present.  When one considers a curved model, the best fit for the curvature
parameter is
$\Omega_{k}^{(0)} = -0.01$ with an uncertainty of few percent \cite{Ade:2013zuv}.

Since the scale factor evolves as a smooth function of time, one is
able to use it as a variable, instead of time, in such a way that $d/dt = a\,  H \, d/da$.  This
change of variable helps to integrate the continuity equation for non-constant $w(a)$ to obtain:
\begin{equation} \label{cont-sol}
\rho(a) = \rho_0 e^{-3 \int [1+w(a)] da/a} .
\end{equation}
If, for instance, one parameterizes dark energy through an analytic function of the scale factor, $w(a)$, one immediately
obtains its solution in terms of
\begin{equation} \label{time-scale-factor}
t = \int \frac{1}{\sqrt{8 \pi G\rho(a)/3}}\frac{da}{a} .
\end{equation}
From Eq. (\ref{time-scale-factor}) one obtains the age of the  Universe in terms of the redshift, $H_0$, and
the density parameters:
\begin{equation}
t_0  = H_{0}^{-1}  \int_{0}^{\infty} \frac{dz}{(1+z) H(z)} .
\label{time_z}
\end{equation}
When combining different cosmological probes one obtains for the  $\Lambda$CDM model an age of
$t_0  =13.81 \pm 0.06$ Gyr \cite{Ade:2013zuv}.

In general, if dark energy is a function  of the redshift, from Eq. (\ref{cont-sol}) one can generalize the
Friedmann equation to:
\begin{equation} \label{hz}
H(z)^2/H_{0}^2  =   \Omega_{M}^{(0)} (1+z)^{3} + \Omega_{\gamma}^{(0)} (1+z)^{4} +
 \Omega_{k}^{(0)} (1+z)^{2} +  \Omega_{DE}^{(0)} {\sf f}(z)  \, ,
\end{equation}
where $DE$ stands for dark energy, and
\begin{equation}
{\sf f}(z) = {\rm exp} \left[ 3 \int_{0}^{z} \frac{1+w(z')}{1+z'}   dz' \right] \, .
\end{equation}
Different DE models can be directly parametrized through $w=w(z)$. The most popular one is perhaps the
Chevalier-Polarski-Linder's \cite{Chevallier:2000qy,Linder:2002et} formula
$w = w_{0} + w_{a} (z/(1+z))$, where $w_{0}$ and $w_{a}$ are constants.

We would like to remark that the first strong evidence for the existence of dark energy, and hence for a
present accelerated expansion of the Universe,
came from fits of supernovae luminosity curves to data \cite{CervantesCota:2011pn}. Two different supernova
groups \cite{Ri98,Ri99,Pe99} found a clear evidence for
$\Lambda$ in the late 90's. The presence of a cosmological constant makes the Universe not only expanding,
but also accelerating and, in addition,
its age is older, and not in conflict with the globular cluster ages \cite{Jimenez:1996at,Richer:2002tg}.
In the course of the years, various supernova
groups have been getting more confident that the data is compatible with the presence of dark energy, dark matter,
and a high value of the Hubble parameter. By moving a little beyond the standard model of cosmology and
letting $w$ be a constant (but
not necessarily $-1$), one of the latest data released, the Union2 compilation \cite{Am10}, reports that the flat
concordance $\Lambda$CDM
model remains an excellent fit to the data, with the best fit to the constant equation-of-state parameter
being $w=-0.997^{+0.050}_{-0.054}$ for a flat Universe, and
$w=-1.035^{+0.055}_{-0.059}$ for a curved Universe. Also, they found that $\Omega_{M}^{(0)}= 0.270 \pm 0.021$
(including baryons and DM)
for fixed $\Omega_{k}^{(0)} =0$. That is, $\Omega_{\Lambda}^{(0)} = 0.730 \pm 0.021$. Using CMB PLANCK data,
these numbers change a few percent,
having little less DE and more DM, as shown by Eq. (\ref{planck_params}).

\subsection{Fluids' chronology}

The standard model of cosmology is described by a set of periods in
which different fluids dominated the dynamics. We first consider a
period of inflation in which the Universe experienced an accelerated
expansion rendering enough {\it e}-folds to explain the horizon and
flatness problems of the old Big Bang theory (see, for instance,
Ref. \cite{CervantesCota:1900zz}). This very early epoch is well
described by an exponential expansion characterized by an equation
of state $w=-1$. This is achieved through a scalar field that slowly
rolls its potential, as we will see in section \ref{inflation}.
Eventually, the scalar field steps down the potential hill and
begins to oscillate, to behave as a fluid of dust ($w=0$)
\cite{Tu83}. This period is thought to be short to let particle
production and to heat the Universe in a period of reheating
\cite{Rh82,Do82,Abb82} and/or preheating \cite{Kof94,Kof96}, for a
modern review see Ref. \cite{Allahverdi:2010xz}. This is needed
since after inflation the Universe is cooled down exponentially and
it is deprived of particles. The new, produced particles,
generically lighter than the scalar field mass, are relativistic ($T
\gg m$, $m$ being its rest mass), and therefore they are well
described by $w =1/3$. This epoch is important because it marks the
beginning of the hot Big Bang theory. In this very early epoch
particle physics theories (such as grand unification schemes) should
describe the details of particle interactions to eventually reach
the lower energies of the well tested standard model of particle
physics. Then, the material content of the Universe consisted of a
hot plasma with photons, protons, neutrons, neutrinos, electrons,
and possibly other particles with very high kinetic energy. After
some cooling of the Universe, some massive particles decayed and
others survived (protons, neutrons, electrons, and DM) whose masses
eventually dominated over the radiation components (photon,
neutrinos, and possibly dark radiation; the latter being any other
relativistic degree of freedom present at that epoch) at the {\it
equality} epoch ($\rho_{\rm rel} = \rho_{m}$) at  $z_{\rm eq} \sim
3402$ \cite{Ade:2013zuv}.  From this epoch and until recent {\it
e}-folds of expansion ($z_{\rm DE} \sim 0.8$), the main matter
component produced effectively no pressure on the expansion and,
therefore, one can accept a model filled with dust, $w =0$, to be
representative for the energy content of the Universe in the
interval $3402 < z < 0.8$. The dust equation of state is then
representative of inert CDM. DM does not (significantly) emit light
and therefore it is dark. Another possibility is that dark matter
interacts weakly, which is generically called WIMP (Weakly
Interacting Massive Particle); the neutralino being the most popular
WIMP candidate. Another popular dark matter candidate is the axion,
a hypothetical particle postulated to explain the conservation of
the CP symmetry in quantum chromodynamics (QCD). Back to the
Universe evolution, from $z \sim 0.8$ \cite{Busca:2012bu} until now
the Universe happens to be accelerating with an equation of state $w
\approx -1$, due to some constant energy that yields a cosmological
constant, $\Lambda =  8 \pi G \rho =$const.  The cosmological
constant is the generic agent of an inflationary solution (see the
$k =0$ solution in Eq. \ref{isc}). The details of the accelerated
expansion are still unknown and it is possible that the expansion is
due to some new fundamental field (e.g., quintessence) that induces
an effective $\Lambda (t) \sim$const. (see section
\ref{dark_energy}). We call (as M. Turner dubbed it) {\it dark
energy} (DE) this new element. Dark energy does not emit light nor
any other particle, and as known so far, it simply behaves as a
(transparent) media that gravitates with an effective negative
pressure. The physics behind dark energy or even the cosmological
constant is unclear since theories of grand unification (or theories
of everything, including gravity) generically predict a vacuum
energy associated with fundamental fields, $<0|T_{\mu \nu}|0>=
<\rho>  g_{\mu \nu}$, that turns out to be very large. This can be
seen by summing the zero-point energies of all normal modes of some
field of mass $m$, to obtain $ <\rho>\approx M^{4}/(16 \pi^{2})$,
where $M$ represents some cut-off in the integration, $M \gg m$.
Then, assuming that GR is valid up to the Planck ($Pl$) scale, one
should take $M\approx 1/\sqrt{8 \pi G}$, which gives $
<\rho>=10^{71}$ GeV$^{4}$. This term plays the role of an effective
cosmological constant $\Lambda = 8 \pi G <\rho> \approx M_{Pl}^{2}
\sim 10^{38}$ GeV$^2$, which must be added to Einstein's Eqs.
(\ref{eesc}), or directly to Eqs. (\ref{frw1sc}) and (\ref{frw2sc}),
yielding an inflationary solution as given by Eq. (\ref{isc}).
However, since the cosmological constant seems to dominate the
dynamics of the Universe nowadays, one has that
\begin{equation}
\Lambda \approx 8 \pi G \rho_{0} = 3 H_{0}^{2} \sim 10^{-83} {\rm GeV}{}^{2},
\label{lsc} \end{equation}
which is very small compared to the value derived above on dimensional
grounds. Thus, the cosmological constraint and the theoretical expectations
are rather dissimilar, by about 121 orders of mag\-ni\-tude!  Even if one
considers symmetries at lower energy scales, the theoretical $\Lambda$ is indeed
smaller, but never as small as the cosmological constraint:
$\Lambda_{GUT}\sim 10^{21}$ GeV$^2$, $\Lambda_{SU(2)}\sim 10^{-29}$ GeV$^2$.
This problem has been reviewed many decades ago \cite{We89,CaPrTu92} and still remains
open.

\section{Scalar fields as perfect fluids} \label{sfbd}

Scalar fields are ubiquitous in cosmology since they allow for modelling different cosmic
dynamics, from inflation \cite{Gu81,Li90} and dark energy \cite{Caldwell:1997ii,CoSaTs06} to dark matter
\cite{Matos:1999et,Magana:2012ph}.
The full characterization of scalar fields is not describable in terms of perfect fluids, but its background dynamics
allows for that. A scalar field with mass, $m_{\phi}$, has an associated Compton wavelength,
$\lambda_{C} = 1/m_{\phi}$. Thus, one
can conceive the fluid picture as a collection of scalar particles with a typical size
of $\lambda_{C}$. For if $\lambda_{C}=H^{-1}_0$ the corresponding
scalar field mass is of course very light, $m_{\phi} =10^{-33}$eV. If $\lambda_{C} < H^{-1}$ the
particle is localizable within the Hubble horizon, otherwise its mass is too light and counts effectively
as a massless particle.

A canonical scalar field ($\phi$) is given by the Lagrangian density
\begin{equation}
{\cal L} =  \frac{1}{2} \,  \partial^{\mu} \phi \, \partial_{\mu} \phi - V(\phi)  \,\ ,
\label{lag-sf}
\end{equation}
where the first term accounts for the kinetic energy and $V(\phi)$ is its potential.

The energy-momentum tensor of the $\phi$-field is
\begin{equation}
T_{\mu \nu} (\phi) = \frac{\partial {\cal L}}{\partial(\partial^{\mu}\phi)}
\partial_{\nu} \phi - {\cal L} g_{\mu \nu} = \partial_{\mu} \phi \partial_{\nu}
\phi - \frac{1}{2} \partial_{\lambda} \phi \partial^{\lambda} \phi  ~
g_{\mu \nu} + V(\phi) g_{\mu \nu} \,\ .
\label{lfin}
\end{equation}
The field energy density and pressure are, by associating
$\rho(\phi)=T_{00}(\phi)$ and $P(\phi)=T_{ii}(\phi)/a^{2}$ (no $i$-sum),
\begin{eqnarray}
\rho(\phi) &=& \frac{1}{2} \dot{\phi}^{2} + V(\phi) + \frac{1}{2a^{2}(t)}
(\nabla\phi)^{2} \approx \frac{1}{2} \dot{\phi}^{2} + V(\phi), \nonumber \\
P(\phi) &=& \frac{1}{2} \dot{\phi}^{2} - V(\phi) - \frac{1}{6a^{2}(t)}
(\nabla\phi)^{2} \approx \frac{1}{2} \dot{\phi}^{2} - V(\phi),
\label{rpin} \end{eqnarray}
where the gradient terms (in comoving coordinates) are neglected. This typically occurs
for the background cosmology and the reason for this is that the Universe is assumed to be sufficiently
homogeneous within a horizon distance.

The equation of state associated to a scalar field is
\begin{equation}
w = \frac{P}{\rho} = \frac{\frac{1}{2} \dot{\phi}^{2} - V(\phi)}{\frac{1}{2} \dot{\phi}^{2} + V(\phi)},
\label{wsf} \end{equation}
with $w$ taking values in the interval $-1 \leq w \leq 1$.

The conservation of energy, Eq. (\ref{frw3asc}), yields, using Eq. (\ref{rpin}), the
equation of motion for the $\phi$-field,
\begin{equation}
\ddot{\phi} + 3 H \dot{\phi} + V'(\phi) = 0 \,\ ,
\label{fin} \end{equation}
where the prime stands for the scalar field derivative. The expansion term plays
the role of a friction, whereas the potential contribution depends upon the scalar field model at hand.

In what follows, we present the main features of two applications of
the scalar field dynamics: inflation and quintessence. We will refer
the reader to recent reviews on these subjects for a more profound
account of these topics, cf.
\cite{Baumann:2009ds,Mazumdar:2010sa,Tsujikawa:2013fta}.

\subsection{Inflation} \label{inflation}

The scalar field responsible for the inflationary dynamics is dubbed the {\it inflaton}. There are
hundreds of models
of inflation and several theoretical aspects related to perturbations (see section \ref{denpertin}),
non-Gaussianities, etc; for a recent review see Ref. \cite{Baumann:2009ds}.
The basics of the dynamics is as follows: the inflaton evolves from an initial value ($\phi_{*}$) down
the hill of the potential, but typically in a {\it slow roll-over} way, to a final state in which
reheating takes place.

In order to get enough {\it e}-folds of inflation the scalar field should stay long time,
compared to the cosmic time, in
a potential `flat region' where the potential is almost constant $V(\phi) \sim V(0)$.
To construct such a flat curvature for the potential and to permit the
$\phi$-field to evolve slowly, one has to impose the slow roll-over
conditions, namely, that $\ddot{\phi}\approx 0$. From Eq. (\ref{fin}), it
implies that  $\dot{\phi} \approx - V^{\prime}/3H$, which in turn means that \cite{StTu84}:
\begin{equation}
\frac{\ddot{\phi}}{3H \dot{\phi}} =
- \frac{V^{\prime\prime}}{9 H^2} + \frac{1}{48 \pi G}
\left(\frac{V^{\prime}}{V}\right)^2 \ll 1 \,\ ,
\label{rolfin}
\end{equation}
or in terms of the dimensionless potential slow-roll parameters,
$\epsilon \equiv 1/(16 \pi G) (V'/V)^{2}$ $ \ll 1$ and $\eta  \equiv 1/(8 \pi G) (V''/V)\ll 1$.

This condition also ensures that
$\rho(\phi) \approx V(\phi)>\frac{1}{2}\dot{\phi}^{2}$, and so from Eq. (\ref{wsf}) one has
$w \approx -1$, which guarantees
an accelerated expansion. However, if the initial conditions are such that at the outset
$\frac{1}{2}\dot{\phi}^{2} \gg  V(\phi)$, then the solution takes the form,
$\dot{\phi}^{2}= {\rm const.}/t^{2}$ and
$\phi = \phi_{0} - A ~ {\rm ln}(1+B t)$, where $A$ and $B$ are constants. Then, the kinetic terms fall faster than
the logarithmic
decrease of a polynomial potential.  Therefore, after some asymptotic time the
Universe will be dominated by its potential and thus, inflation follows
\cite{Li90}. However, in other gravity theories the kinetic terms play an important role and
could prevent the Universe from inflation \cite{CervantesCota:1994zf,CervantesCota:1995tz}.

The scalar field solution with $w \approx -1$, considered in  Eq. (\ref{ecsc}), emulates a vacuum energy term
or a cosmological constant.  Given the slow roll-over of the $\phi$-field this behaviour happens for a minimum of
$N$ {\it e}-folds of expansion in which the Hubble rate is effectively given by
\begin{equation}
H^{2} = \frac{8 \pi G}{3} V(\phi\approx {\rm const.}) \,\ .
\label{cpin} \end{equation}
In this way, $H \approx$const. and the scale factor exhibits an exponential behaviour, as given by Eq. (\ref{isc}).
Strictly speaking, during inflation $\phi$ is an increasing function of time,
since $V'<0$ in Eq. (\ref{fin}).  However, under slow roll-over conditions its
characteristic evolution time will be much greater than the cosmological time.  Therefore, $H$
will be a very slow, monotonically decreasing function of time.

\begin{figure}[ht]
\includegraphics[width=12cm]{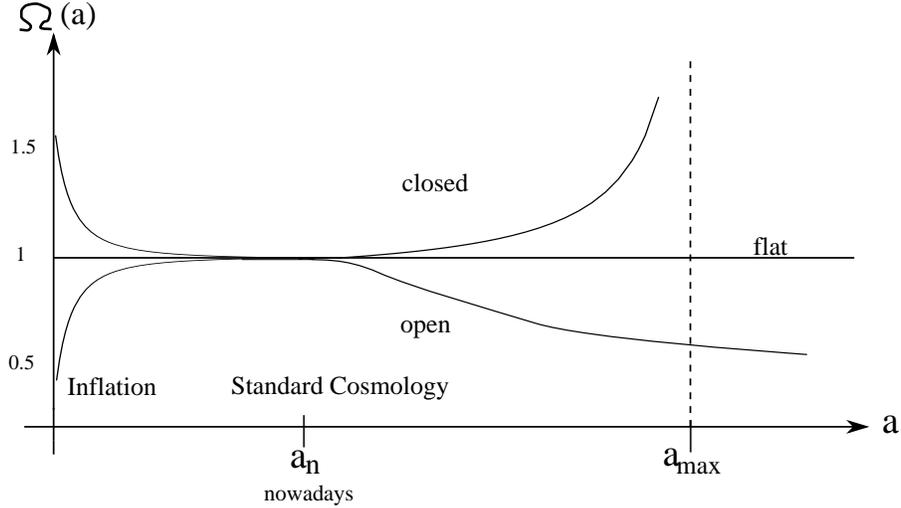}
\caption[The density parameter as a function of the scale factor]{The
parameter $\Omega$ as a function of the scale factor, $a$, during
inflation and thereafter in a  radiation/matter dominated Universe.  Inflation makes
the space to look like as flat, even if it is initially curved. If there are enough {\it e}-folds of inflation to
solve the horizon problem, it implies that the Universe nowadays is still flat. Later on, the behaviour
is as in Fig. \ref{omegasc}.
\label{omein}}
\end{figure}

Inflation lasts for a sufficient number of $N$ {\it e}-folds to solve the horizon and flatness problems
in cosmology, and this
depends very much on the energy scale of inflation. In standard inflationary scenarios $N\sim 60$. This ensures
that a possible curved model will look like a flat one for all the expansion history, including today (see
Fig. \ref{omein} and compare with Fig. \ref{omegasc}).

Among the multiple inflaton potentials considered in the literature, the most favoured models by the PLANCK CMBR
temperature map fits \cite{Ade:2013uln}
are those having potentials with $V''< 0$. Exponential potential models, the simplest hybrid inflationary
models, and monomial potential models of degree
$n \ge 2$ do not provide a good fit to the data. The most favoured models are Hill-top models, a simple symmetry
breaking potential,
natural inflation, $R^2$ inflation, and non-minimal coupled to gravity with a Mexican-hat potential; see
Ref. \cite{Ade:2013uln} for details.

\subsection{Dark energy: quintessence} \label{dark_energy}

Dark energy is a generic name for an energetic ``fluid'' that has had little or no evolution in the past
few giga-years of the cosmic expansion. Since then
dark energy dominates the total density of the Universe over all other components (dark matter, baryons,
photons, and neutrinos).
During dark energy domination, the Hubble parameter, as given by Eq. (\ref{frw1sc}), is basically a constant.
Thus, a cosmological constant added to the
gravitational theory is the simplest candidate for dark energy that fits the data from the different cosmic probes.
There are at least seven independent observations that imply the presence of dark energy: the ages of some
globular clusters surpasses the age of the Universe in models without dark energy \cite{Jimenez:1996at,Richer:2002tg};
the supernovae best fits
to distance moduli \cite{Ri98,Ri99,Pe99}; the dynamics of clusters of galaxies \cite{Allen:2004cd}; the
combination of the CMBR lensing deflection power
spectrum  with temperature and polarization power spectra \cite{Sherwin:2011gv}; the measurements of the
integrated Sachs-Wolf
effect \cite{Giannantonio:2012aa}; the measurements of Baryon Acoustic Oscillations (BAO) \cite{Eisenstein:2005su};
and the change of the Hubble rate behaviour
from galaxy surveys \cite{Busca:2012bu}.

Another possible candidate for dark energy is a canonical scalar field, dubbed quintessence \cite{Caldwell:1997ii}.
The equations governing the scalar field dynamics in a cosmological background
are those displayed in section \ref{sfbd}. Basically, the FRW equations, i.e., Eqs. (\ref{frw1sc}) and
(\ref{frw2sc}), are now fulfilled with the density and pressure terms given by Eqs. (\ref{rpin}).
To complete the whole picture, we add the rest of the known four material elements (dark matter, baryons,
photons, and neutrinos) to the scalar field.

In a similar fashion to inflation, one demands that $V(\phi)>\frac{1}{2}\dot{\phi}^{2}$ has a flat potential
and allows for an accelerated behaviour. One may
again use the slow roll-over parameters ($\epsilon, \eta$) to ensure an accelerated dynamics, but here we have
the other four components that may spoil the exact accelerated
dynamics. Still, this approach works well.

Originally, runaway potentials were considered, but nowadays there is a vast set of models that achieve the desired
accelerated dynamics, including non-standard kinetic terms \cite{CoSaTs06,DeSantiago:2012nk} or scalar fields
interacting with
matter \cite{Aviles:2010ui}, among many others. To avoid the over dominance of the scalar field during the
early stages of the cosmic dynamics, one
looks for scaling properties (of tracker nature) of the scalar field dynamics in which the field energy
density ($\rho_{\phi}$) evolves proportionally to the
material fluid energy density ($\rho_m$) with $\rho_{\phi} < \rho_m$, and only until recently the scalar
field turns to dominate. Depending on the evolution of the scalar-field equation of state, Eq. (\ref{wsf}),
quintessence models can be freezing or thawing \cite{Caldwell:2005tm}. The former class is when the
scalar field gradually slows down to
eventually freeze in a constant value.
The latter class implies that the scalar field has recently started to change from a past constant value.
These behaviours can in principle be
tested (see Ref. \cite{Tsujikawa:2013fta} for a recent review on the subject).

\section{Thermodynamics in the early Universe}

In the early Universe one considers a plasma of particles and
their antiparticles, as was done originally by
Gamow \cite{Ga46}, who first considered a physical scenario for the hot Big Bang model as a
description of the beginning of the Universe. Later on, with the development of modern particle
physics theories in the 70's it was unavoidable to think about a physical
scenario which should include the ``new'' physics for the early Universe.
It was also realized that the physics described by GR should not be applied
beyond Planckian initial conditions, because there the quantum corrections to
the metric tensor become very important, a theory which is still in progress.

After preheating/reheating, one assumes that the Universe is filled with a plasma of relativistic particles
which include quarks, leptons, and gauge and Higgs bosons, all in thermal
equilibrium at a very high temperature, $T$, with some gauge symmetry
dictated by a particle physics theory.

Theoretically,  one introduces some thermodynamic
consi\-de\-rations necessary for the description of the physical content of
the Universe, which we would like to present here.  Assuming an ideal-gas
approximation, the number density $n_{i}$ of particles of type $i$, with a
momentum $p$, is given by a Fermi or Bose distribution \cite{KoTu90}:

\begin{equation}
n_{i} = \frac{g_{i}}{(2 \pi)^{3}}
\int \frac{d^{3}p}{e^{(E_{i}- \mu_{i})/T} \pm 1} \,\ ,
\label{dissc}
\end{equation}
where $E_{i}=\sqrt{m_{i}^{2}+ p^{2}}$ is the particle energy, $\mu_{i}$ is
the chemical potential, the sign $(+)$ applies for fermions and $(-)$ for
bosons, and  $g_{i}$ is the number of spin states. One has that
$g_{i}=2$ for photons, quarks, baryons, electrons, muons, taus, and their
antiparticles, but $g_{i}=1$ for neutrinos because they are only left-handed.
For the particles existing in the early Universe one usually assumes
that $\mu_{i} =0$: one expects that in any particle reaction the $\mu_{i}$
are conserved, just as the charge, energy, spin, and lepton and baryon number
are. For a photon, which can be created and/or annihilated after some
particle's collisions, its number density, $n_{\gamma}$, must not be conserved
and its distribution with $\mu_{\gamma}=0, \,\ E = p = \omega$, reduces to the
Planckian one.  For other constituents, in order to determine
the $\mu_{i}$, one needs
$n_{i}$. Note from Eq. (\ref{dissc}) that for large $\mu_{i}>0, \,\  n_{i}$ is large
too. One does not know $n_{i}$ in advance. However, the WMAP data constrains the baryon density
at nucleosynthesis such that
\cite{CyFiOlSk04}:
\begin{equation}
\eta \equiv \frac{n_{B}}{n_{\gamma}} \equiv
\frac{n_{\rm baryons} - n_{\rm anti-baryons}}{n_{\gamma}} =
6.14 \pm 0.25 \times 10^{-10} \,\ .
\label{barasc}
\end{equation}
The smallness of the baryon number density, $n_{B}$, relative to the
photon's, suggests that $n_{\rm leptons}$ may also be small compared to
$n_{\gamma }$. Therefore, one takes for granted that
$\mu_{i}=0$ for all particles. The ratio  $n_{B}/n_{\gamma}$ is very small, but not zero. The reason of
why matter prevailed over antimatter is one of the puzzles of the
standard model of cosmology called {\it baryogenesis} \cite{KoTu90}. There are some attempts
to achieve baryogenesis at low energy scales, as low as few GeV or TeV
\cite{CervantesCota:1995tz,low-scale-baryo90s,Coh93,Tro99,Bez08}.
Recent attempts to solve this problem are looking for prior to lepton asymmetry, {\it leptogenesis},
generated in the decay of a heavy sterile neutrino \cite{DaNaNi08}, to then end with baryogenesis.

The above approximation allows one to treat the density and pressure of all
particles as a function of the temperature only. According to the second law
of thermodynamics, one has \cite{We72}:
\begin{equation}
d S(V,T) = \frac{1}{T}[d(\rho V) + P dV] ,
\end{equation}
where $S$ is the entropy in a volume $V\sim a^{3}(t)$, with
$\rho = \rho(T)$ and $P = P(T)$ in equilibrium.
Furthermore, the following integrability condition
$\frac{\partial^{2} S}{\partial T \partial V} =
 \frac{\partial^{2} S}{\partial V \partial T}$  is also valid, which turns
out to be
\begin{equation}
\frac{d P}{dT} = \frac{ \rho  + P }{T}   \,\ .
\label{psc}
\end{equation}
On the other hand, the energy conservation law, Eq. (\ref{frw3asc}), leads to
\begin{equation}
\frac{d}{dt}\left[\frac{a^{3}(t)}{T}(\rho  + P)\right]=0,
\label{cosc}
\end{equation}
after using Eq. (\ref{psc}).
Using Eq. (\ref{psc}) again, the entropy equation can be written as
$
d S(V,T) = \frac{1}{T}d[(\rho+P) V] - \frac{V}{T^{2}}(\rho+P) dT
$. These last two equations imply that the entropy is a constant of motion:
\begin{equation}
S = \frac{a^{3}}{T}[\rho  + P ] = {\rm const.}  \, .
\label{entconsc}
\end{equation}

Moreover, the density and pressure are given by
\begin{equation}
\rho \equiv  \int E_{i} n_{i} dp \,\ , \qquad
P  \equiv   \int \frac{p^{2}}{3 E_{i}}  n_{i} dp \,\ .
\label{rpsc}\end{equation}
For photons or ultra-relativistic fluids, $E=p$, and the above equations become
$P=\frac{1}{3} \rho $, thus confirming Eq. (\ref{ecsc}) for $w = 1/3$.
After integration of Eq. (\ref{psc}), it comes out that
\begin{equation}
\rho = b T^{4} \, ,
\label{bbsc}\end{equation}
where $b$ is a constant of integration. In the real Universe there are many
relativistic particles present, each of which contributes like Eq.
(\ref{bbsc}). By including all of them, $\rho = \sum_{i} \rho_{i}$ and
$P = \sum_{i} P_{i}$, where the summations are over all relativistic species, one has that
$b(T)=\frac{\pi^{2}}{30} (N_{B} + \frac{7}{8}N_{F} )$,
which depends on the effective relativistic degrees of freedom of
bosons ($N_{B}$) and fermions ($N_{F}$). Therefore, this quantity varies with
the temperature. Different $i$-species remain relativistic until some
characteristic
temperature $T\approx m_{i}$ and after this point $N_{F_{i}}$
(or $N_{B_{i}}$) no longer contributes to $b(T)$. The factor 7/8 accounts for
the different statistics of the particles [see
Eq. (\ref{dissc})]. In the standard model of particle physics
$b\approx 1$ for $T \ll 1$ MeV and $b\approx 35$ for $T>300$ GeV \cite{KoTu90}.
In particular, one accounts for the effective number of neutrinos ($N_{\rm eff}$) in terms of photons'
degrees of freedom as
\begin{equation}
\frac{\rho_{\nu}}{\rho_{\gamma}} =  \frac{7}{8} \left( \frac{4}{11} \right)^{4/3} \: N_{\rm eff} ,
\label{neff}
\end{equation}
with $N_{\rm eff} =3.046$ for standard model neutrino species \cite{Mangano:2005cc}.  Extra neutrino-type
relativistic species -- dark radiation -- should augment $N_{\rm eff}$, as was recently suggested
from measurements
of different cosmological probes. Combining PLANCK with previous CMB data and Hubble Space Telescope
measurements, it has been concluded
that $N_{\rm eff}=3.6 \pm0.5$ with a $95\%$ confidence level \cite{DiValentino:2013qma}.

For  relativistic particles, we obtain from Eq. (\ref{dissc}) that
\begin{equation}
n = c T^{3}, \,\ \,\ {\rm with} \,\
 c=\frac{\zeta(3)}{\pi^{2}} (N_{B} + \frac{3}{4} N_{F}) \,\ .
\label{nsc} \end{equation}
where $\zeta(3)\approx 1.2$ is the Riemann zeta function of 3. Nowadays,
$n_{\gamma} \approx 411 T^{3}_{2.73}$ cm$^{-3}$, where
$T_{2.73}\equiv T_{\gamma_{0}}/(2.73 {\rm K})$.
The precise measured value is  $T_{\gamma_{0}} = 2.72548 \pm 0.00057 {}^{\circ}\!K$ \cite{Fixsen:2009ug}.
The mean energy per photon is $6.34 \times 10^{-4}$ eV which corresponds to a wavelength of 2 millimetres,
and hence it is called cosmic ``microwave'' background radiation.

Using the relativistic equation of state given above ($w=1/3$), From Eq. (\ref{entconsc}) it
follows that $T\sim 1/a(t)$. From its solution in Eq.
(\ref{frwsolsc}) one has
\begin{equation}
T = \sqrt[4]{\frac{M_{\frac{1}{3}}}{b}} \frac{1}{a(t)}
  = \sqrt[4]{\frac{3}{32 \pi G b}} \frac{1}{(t-t_{*})^{\frac{1}{2}}} \,\ ,
\label{tradsc}
\end{equation}
which predicts a decreasing temperature behaviour as the Universe expands. Then,
initially at the Big Bang, $t=t_{*}$ implies that $T_{*}= \infty$, and so the
Universe was not only very dense but also very hot. As time evolves the Universe expands,
cools down, and its density diminishes.

The entropy for an effective relativistic fluid is given by Eq. (\ref{entconsc})
together with its equation of state and Eq. (\ref{bbsc}), i.e.,
$S=\frac{4}{3} ~ b ~ (a ~T)^{3}= {\rm const.} $
Combining this with Eq. (\ref{tradsc}), one can compute the value of
$M_{\frac{1}{3}}$ to be
$M_{\frac{1}{3}}=(\frac{3}{4}S)^{4/3} /b^{1/3}$
$\approx 10^{116}$, since $b\approx 35$ and the photon entropy
$S_{0} = \frac{4}{3} ~ b ~ (a_{0} ~T_{0})^{3} \approx 10^{88}$ for
$a_{0} \rightarrow d_{H}(t_{0})=10^{28}$ cm and
$T_{\gamma_{0}}=2.73$ K, as evaluated at the present time. One defines the
entropy per unit volume, {\it entropy density}, to be
$s \equiv S/V = \frac{4}{3} \frac{\pi^{2}}{30} (N_{B}+\frac{7}{8}N_{F}) T^{3}$,
then at the present time $s \approx 7 n_{\gamma}$. The nucleosynthesis bound on $\eta$,
Eq. (\ref{barasc}), implies that $n_{B}/s \approx  10^{-11}$.

We now consider particles in their non-relativistic limit ($m \gg T$). From
Eq. (\ref{dissc}) one obtains for both bosons and fermions that
\begin{equation}
n= g  \left(\frac{m T}{2 \pi}\right)^{3/2} e^{-m/T} \,\ .
\label{nnrsc} \end{equation}
The abundance of equilibrium massive particles decreases exponentially once
they become non-relativistic. This situation is referred to as {\it in equilibrium
annihilation}. Their density and pressure are given through
Eqs. (\ref{rpsc}) and (\ref{nnrsc}) by
$\rho =n m$ and $ P  = n T \ll \rho$.
Therefore, using these last two equations, the entropy for non-relativistic particles, given by Eq. (\ref{entconsc}),
diminishes also exponentially during the in equilibrium annihilation.
The entropy of these particles is transferred to that of
the relativistic components by augmenting their temperature. Hence,
the constant total entropy is essentially the same as the one given above, but the $i$-species
contributing to it are just those
which are in equilibrium and maintain their relativistic behaviour, that is,
particles without mass such as photons.

\begin{figure}[ht]
\includegraphics[width=10cm]{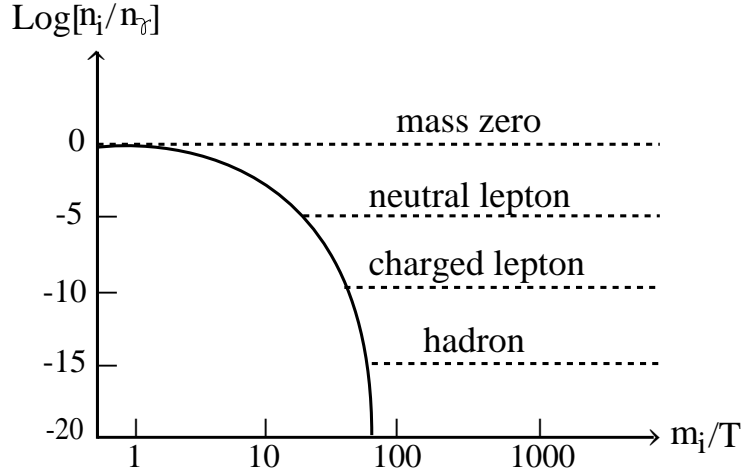}
\vskip -3.5cm
\caption[Evolution of the particle density of different $i-$species]{Evolution
of the particle density for different $i$-species. If a given
$i$-species is in equilibrium, its abundance diminishes exponentially after the
particle becomes non-relativistic (solid line). However, interactions of
an $i$-species can freeze out, then it decouples
from equilibrium and maintains its abundance (dashed line).
Figure adap\-ted from Ref. \cite{KoTu90}.
\label{fresc} }
\end{figure}

Having introduced the abundances of the different particle types, we
would like to comment on the equilibrium conditions for the
constituents of the Universe as it evolves. This is especially important
in order to have an idea of whether or not a given $i$-species disappears or
decouples from the primordial brew. To see this, let us consider
$n_{i}$ when the Universe
temperature, $T$, is such that (a) $T \gg m_{i}$, during the ultra-relativistic
stage of some particles of type $i$ and (b) $T \ll m_{i}$, when the particles
$i$ are non-relativistic, both cases in thermal equilibrium.  From Eq.
(\ref{nsc}), one has that for the former case $n_{i} \sim T^{3}$ and the total
number of particles, $\sim n_{i} a^{3}$, remains constant,
whereas for the latter case, using Eq. (\ref{nnrsc}),
$n_{i} \sim T^{3/2} e^{-m_{i}/T}$, i.e.,
when the Universe temperature goes down below $m_{i}$, the number density of
the $i$-species significantly diminishes; it occurs an in equilibrium
annihilation. Let us take as an example the neutron-proton
annihilation. Then we have
\begin{equation}
\frac{n_{n}}{n_{p}} \sim \exp\left(\frac{m_{p}-m_{n}}{T}\right)
= \exp\left(-\frac{1.5 \times 10^{10} {\rm K}}{T}\right) ,
\label{npsc}\end{equation}
which drops with the temperature from near 1 at
$T \ge 10^{12}$ K to about 5/6 at
$T \approx 10^{11}$ K and 3/5 at
$T \approx 3 \times 10^{10}$ K \cite{Na02}. If this is valid forever,
we then end up without massive particles and our Universe would have been
consisted only of radiative components. However, our own existence prevents that!
Therefore, eventually the in equilibrium annihilation had to be stopped. The
quest is now to freeze out this ratio to $n_{n}/n_{p} \approx 1/6$ (due to
neutron decays) until the time when nucleosynthesis begins (i.e., when $n_{n}/n_{p}$
reduces to 1/7) in order to leave the correct number of hadrons and achieve later
successful nucleosynthesis. The answer comes from comparing the
Universe expansion rate, $H$, with the particle physics
reaction rates, $\Gamma$. Hence, for $H<\Gamma$ the particles interact
with each other faster than the Universe expansion rate, then equilibrium is
established. For $H>\Gamma$ the particles cease
to interact effectively, then thermal equilibrium drops out. This is only approximately
true; a proper account of that involves a Boltzmann equation analysis.
For that analysis numerical integration should be carried out
in which annihilation rates are balanced
with inverse processes, see for example Ref. \cite{St79,KoTu90}. In this way,
the more interacting the particles are, the longer they remain
in equilibrium annihilation and, therefore, the lower their number
densities are after some time, e.g., baryons vanish first, then charged
leptons, neutral leptons, etc.;
finally, the massless photons and neutrinos, whose particle numbers
remain constant, as it was mentioned above (see Fig. \ref{fresc}).
Note that if interactions of a given $i$-species freeze out while it is still
relativistic, then its abundance will be significant at the present time and will account for
dark radiation, as was recently suggested in Ref. \cite{DiValentino:2013qma}.

It is worth mentioning that if the Universe would expand faster, then the
temperature of decoupling, when $H \sim \Gamma$, would be higher and thus, the fixed
ratio $n_{n}/n_{p}$ would be greater and the $^{4}$He abundance would be higher,
leading to profound implications in the nucleosynthesis
of light elements. Thus, the expansion rate cannot be
arbitrarily modified during the equilibrium era of some particles. Furthermore, if a particle
species is still highly relativistic ($T \gg m_{i}$) or highly non-relativistic
($T \ll m_{i}$), when decoupling from the primordial plasma occurs, it maintains
an equilibrium distribution; the former being characterized by $T_{r} a = $const. and
the latter by $T_{m} a^{2} = $const. [cf. Eq. (\ref{tmsc})].

There are also some other examples of decoupling, such as  neutrino decoupling:
during nucleosynthesis there exist reactions, e.g.
$ \nu \bar{\nu} \longleftrightarrow e^{+} e^{-}$, which maintain neutrinos
efficiently coupled to the original plasma ($\Gamma>H$) until about 1 MeV,
since $\Gamma /H \approx T^{3}$ MeV$^{-3}$. The reactions are no longer efficient below
1 MeV and therefore neutrinos decouple and continue
evolving with a temperature $T_{\nu}\sim 1/a$. Then, at
$T \, {}^{>}_{\sim} \, m_{e}=0.51$MeV the particles in equilibrium are photons
(with $N_{B}=2$) and electron-positron pairs (with $N_{F}=4$), which contribute
to the entropy with
$b(T)=(11/2)(\pi^{2}/30)$. Later, when the temperature drops to
$T \ll m_{e}$, the reactions are again no longer efficient ($\Gamma < H$) and,
after the $e^{\pm}$ pair annihilation, there will be
only photons in equilibrium with $b(T)=2(\pi^{2}/30)$. Since the
total entropy, $S = (4/3) b (aT)^{3}$, must be conserved, a decrease of
$b(T)$ must be balanced with an increase of the radiation temperature so
that $T_{\gamma}/T_{\nu} = \left(11/4\right)^{1/3}$,
which should remain so until today, implying the existence of a cosmic
background of neutrinos with a present temperature of
$T_{\nu_{0}} = 1.95$ K. This cosmic relic has not been measured yet.

Another example is the gravitation decoupling, which should be also
present if gravitons were in thermal equilibrium at the Planck time and then
decouple. Today, the temperature background should be characterized at most
by $T_{\rm grav} = \left(4/107\right)^{1/3}$ K $\approx 0.91$ K.

For the matter dominated era we have stressed that effectively one has $P=0$. Next we
will see the reason for this. First, consider an ideal gas (such as atomic
hydrogen) with mass $m$, then $\rho = n m + \frac{3}{2} n  T_{m}$ and
$P=n  T_{m}$.  From  Eq. (\ref{cosc}), one equivalently obtains that
\begin{equation}
\frac{d}{d a} (\rho a^{3}(t)) = -3 P \, a^{2}(t),
\label{co1sc}
\end{equation}
which after substitution of $\rho $ and $P$, as given above, becomes
\begin{equation}
\frac{d}{d a} \left( n m a^{3}(t) + \frac{3}{2} n \, T_{m} a^{3}(t) \right) =
-3 \, n \, T_{m} a^{2}(t),
\label{co2sc}
\end{equation}
where $n m a^{3}(t)$ is a constant. This equation yields
\begin{equation}
T_{m} a^{2}(t)= {\rm const.} \,\ ,
\label{tmsc}
\end{equation}
so that the matter temperature drops faster than the radiation temperature
as the Universe expands [cf. Eq. (\ref{tradsc})]. Now, if one
considers both radiation and matter, one has that
$\rho = n m + \frac{3}{2} n  T_{m} + b T_{r}^{4}$ and
$P=n  T_{m} + \frac{1}{3} b T_{r}^{4}$. The source of the Universe expansion
is proportional to
$\rho +3 P =  n m + \frac{9}{2} n  T_{m} + 2  b T_{r}^{4}$, where
the first term dominates over the second, precisely because $T_{m}$ decreases
very rapidly. The third term diminishes as $\sim 1/a^{4}$, whereas the first does it as
$\sim 1/a^{3}$. After the time of density equalization,
$\rho_{m} = \rho_{r}$,
the matter density term is greater than the others and this explains why one assumes a
zero pressure for that era.

From now on, when we refer to the temperature, $T$, it should be related to
the radiation temperature. The detailed description of the Universe thermal evolution for the different
particle types, depending on their masses, cross-sections, etc.,
is well described in many textbooks, going from the physics
known in the early 70's \cite{We72} to
the late 80's \cite{KoTu90}, and therefore it will not be presented here.
However, we notice that as the Universe cools down a series of
spontaneous symmetry--breaking (SSB) phase transitions are expected to occur.
The type and/or nature of these transitions depend on the specific particle
physics theory considered. Among the most popular ones are the Grand
Unification Theories (GUT's), which bring together all known interactions
except for gravity.  One could also regard the
standard model of particle physics or some extensions of it. Ultimately,
when constructing a cosmological theory, one should settle the energy
scale that one wants to describe physically. For instance, at a temperature between
$10^{14}$ GeV and $10^{16}$ GeV a transition to the $SU(5)$ GUT should take
place, if this theory would be valid, in which a Higgs field breaks this symmetry
to $SU(3)_{C} \times SU(2)_{W} \times U(1)_{HC}$, a process through which
some bosons acquire their masses. Due to the gauge symmetry, there
are color (C), weak (W), and hypercharge (HC) conservation, as the subscripts
indicate.  Later on, when the Universe evolved to around 150 GeV the
electroweak phase transition took place in which the standard model Higgs field broke
the symmetry $SU(3)_{C} \times SU(2)_{W} \times U(1)_{HC}$ to
$SU(3)_{C} \times U(1)_{EM}$; through this breaking fermions
acquired their masses.  At this stage, there were only color and electromagnetic
(EM) charge conservation, due to the gauge symmetry.
Afterwards, around a temperature of 200 MeV \cite{Aoki:2009sc} the Universe should undergo
a transition associated to the chiral symmetry-breaking and color confinement from which baryons
and mesons were formed out of quarks. Subsequently, at approximately
10 MeV \cite{Dolgov:2002cv} the synthesis of light elements (nucleosynthesis) began and lasted until temperatures
below 100 keV, when most of the today observed hydrogen, helium, and some other light elements
abundances were produced. So far the nucleosynthesis represents the earliest scenario
tested in the standard model of cosmology. After some thousands of years ($z \sim 3402$ \cite{Ade:2013zuv}),
the Universe became matter dominated, over the radiation components.
At about 380,000 years ($z \sim 1090$ \cite{wmap7,Ade:2013zuv})
recombination took place, that is, the hydrogen ions and electrons combined to form neutral hydrogen atoms,
then matter and electromagnetic radiation decoupled from each other. At this moment, the (baryonic) matter
structure began to form. Since that moment,
the surface of last scattering of the CMBR evolved as an imprint of the early
Universe. This is the light that Penzias and Wilson \cite{Penzias:1965wn} first measured, and that, later on, was
measured in more
detail by BOOMERANG\cite{deB00}, MAXIMA \cite{Ha00}, COBE \cite{Sm92}, WMAP \cite{Bennett:2012fp}, and
now PLANCK \cite{Ade:2013zuv}, among other probes.

\section{Perturbed fluids in the Universe}

In the previous sections, we have outlined how the evolution of a homogeneous Universe can be described by
means of few equations and simple concepts such as the ideal perfect fluids.
The next step is that of introducing in this scenario small inhomogeneities
that can be treated as first order perturbations to those equations,
the goal being the description of the structures we see today in the Universe.
This perturbative approach is sufficient to accurately explain the small temperature anisotropies
($\Delta T/T\sim10^{-5}$) observed
in the CMBR today,
but it can only describe the distribution of matter today at those scales that are still in the linear regime.
At the present epoch, scales smaller than $\sim 30\;{\rm Mpc}\; h^{-1}$ \cite{reid10} have
already entered the non linear-regime ($\Delta\rho/\rho>>1$)
due to the fact that matter tends to cluster under the effects of gravity.
These scales can therefore be described only by means of numerical or semi-numerical approaches
\cite{carlson09}.

The approach is quite straightforward. It involves a differential equation for the density perturbation
of each individual constituent: scalar fields in inflation, or baryons, radiation, neutrinos, DM, and
DE (usually treated as cosmological constant) in later times, and in general it needs to be solved numerically.
In the context of the metric theories of gravity, and in particular GR, the metric is treated as
the general expansion term
$g_{\mu\nu}^{(0)}$ plus a perturbation $h_{\mu\nu}$:
\begin{equation}
\label{hmn}
g_{\mu\nu}=g_{\mu\nu}^{(0)}+h_{\mu\nu},
\end{equation}
with $h_{\mu\nu}<<g_{\mu\nu}^{(0)}$, where $^{(0)}$ indicates unperturbed homogeneous quantities.

Inhomogeneities in the distribution of the components of the Universe are a source of scalar perturbations
of the metric. Nevertheless, vector or tensor perturbations can modify the metric as well. The standard
cosmological model does not predict vector perturbations that would introduce off-diagonal terms in the metric
tensor. These
perturbations would produce vortex motions in the primordial plasma, which are expected to rapidly decay.
Models with topological
defects or inhomogeneous primordial magnetic fields instead predict a consistent fraction of vector
perturbations \cite{seljak97b,turok97,kim09}.

On the other hand, the standard cosmological model predicts the production of gravitational waves during
the epoch of inflation, when the Universe
expanded exponentially. Gravitational waves induce tensor
perturbations  $h^T_{\mu\nu}$  on the metric of the type:
\begin{displaymath}
h^T_{\mu\nu} = a^2
\left( \begin{array}{cccc}
0 & 0 & 0 & 0 \\
0& h_+ & h_{\times} & 0 \\
0& h_{\times} & -h_+ & 0 \\
0 & 0 & 0 & 0
\end{array} \right)
\end{displaymath}
where $h_+$ and $h_{\times}$  are the polarization directions of the gravitational wave. This tensor is
traceless, symmetric,
and divergentless, i.e. it perturbs the time space orthogonally to the direction of propagation of the wave.
The amplitudes of these
tensor perturbations are expected to be small compared to the scalar ones, and therefore negligible in a
first approximation
as far as we are interested in studying the perturbations of the metric tensor. Nevertheless, these waves
are expected to leave
an imprint in the polarization of the CMBR, and their eventual detection would unveil an extremely rich
source of information about an epoch of the Universe that is very hardly observable otherwise.

It is important to underline that choosing to model the metric perturbations corresponds to choosing
a \emph{gauge}, i.e. a specific
coordinate system in which the metric tensor is represented. Changing the coordinate system, of course,
do not change the physics, but can
remarkably vary the difficulty of the calculations and ease the understanding of the physical meaning of the
different quantities.
In order to solve the perturbed equations one chooses convenient gauges for the different expansion epochs and
depending on whether the formalism is theoretical or numerical, as we will see below.

The presence of weak inhomogeneous gravitational fields introduces small perturbations in the metric tensor.
The most general perturbation to the FRW metric is:
 \begin{equation} \label{gen-pert}
  ds^2= a^{2} (\eta)\,  \left[ -(1+2 A) \, d\eta^{2} - B_{i} dx^{i} d \eta +
[(1+2 D) \delta_{ij} + 2 E_{ij}] dx^{i} dx^{j} \right] ,
\end{equation}
where $\eta$ and $x^i$ are comoving coordinates in which the expansion factor $a(\eta)$ is factored
out. Different choices of them imply different gauges. We refer to Refs. \cite{MuFeBr92,ma94,ma95,LyLi09}
for an account of the physical
meaning of the metric potentials and a full treatment of the perturbations.

In correspondence to the above metric perturbations, the energy-momentum tensor is also perturbed. One has:
\begin{eqnarray} \label{T_pert}
 T^{0}_{0} & = & - (\rho + \delta \rho),  \nonumber \\
 T^{0}_{i} & = & (\rho + P) (v_{i} -B_{i}), \nonumber \\
 T^{i}_{0} & = & -(\rho + P) v^{i}, \nonumber \\
 T^{i}_{j} & = & (P + \delta P) \delta^{i}_{j} + \pi^{i}_{j},
 \end{eqnarray}
where $v^{i} = dr^{i} /dt$ is the velocity in local orthonormal coordinates
[$dt = a\ (1+A) \, d\eta; ~ dr^{i} = a\,  dx^{i}$] and
$\pi^{i}_{j}$ are the anisotropic stresses; if they are null the perturbed fluid is also a perfect fluid.
Anisotropic stresses
are important before last scattering, when the primordial plasma was coupled. Later on, when structure
formation begins they are set to zero.

A convenient gauge choice is given through two scalar functions $\Phi(\eta, x^{i})$ and $\Psi(\eta, x^{i})$
as \cite{MuFeBr92}:
\begin{equation}
 ds^2= a^{2} (\eta)\,  \left[ -[1+2\Phi(\eta, x^{i})] \, d\eta^2+ [1+2 \Psi(\eta, x^{i})] dx_{i} dx^{i} \right] ,
\end{equation}
where the perturbed part of the metric tensor is:
\begin{equation}
 h_{00}(\eta, x^{i}) = -2\Phi(\eta, x^{i}), \quad h_{0i}(\eta, x^{i}) = 0, \quad
h_{ij}(\eta, x^{i}) = a^2\delta_{ij}(2\Psi(\eta, x^{i})) .
\end{equation}

This metric is just a generalization of the well-known metric for a weak gravitational field usually
described in the textbooks (e.g. chapter 18
of Ref. \cite{MiThWh73}) for the case of a static Universe [$a(\eta)=1$]. The function $\Phi$ describes
Newton's gravitational field, while
$\Psi$ is the perturbation of the space curvature. The above gauge is the \emph{Newtonian conformal gauge},
which has the advantage
of having a diagonal metric tensor $g_{\mu\nu}$ in which the coordinates are totally fixed with no
residual gauge modes and therefore
with a straightforward interpretation of the functions introduced.

Another example of a gauge that is
particularly popular in the literature is the \emph{synchronous gauge}, defined by:
\begin{equation}
 ds^2=a^2 (\eta) [-d\eta^2+(\delta_{i,j}+h_{i,j}) \, dx^{i} \, dx^{j}] ,
\end{equation}
which is especially used in numerical codes for calculations of the anisotropies and inhomogeneities in
the Universe. It behaves well numerically by choosing
that observers fall freely without changing their spatial coordinates.

The full perturbed equations are obtained by substituting the above expressions, for the chosen gauge, into the
Einstein equations. Alternatively, one may obtain the continuity equation from the time ($\mu=0$) component of
Eq. (\ref{cons-law1}) and the Euler equation from its space sector ($\mu=i$). Here we do not write down
the perturbed equations for any particular gauge, but rather refer the reader to standard textbooks
\cite{LyLi09}, where these equations are fully described.

\subsection{Perturbations during inflation \label{denpertin}}

The primeval fluctuations are thought to be present at the very beginning of time, at the inflationary epoch.
The perturbations are produced by quantum fluctuations of the $\phi$-field during the accelerated stage.
These fluctuations are usually
studied in the {\it comoving gauge} in which the scalar field is equal to its perturbed value at any given
time during inflation
and therefore, the perturbation information resides in the metric components (see
Refs. \cite{MuFeBr92,CervantesCota:1900zz,LyLi09} for reviews on the subject).

To understand how perturbations evolve it is necessary to introduce the concept of horizon
\cite{CervantesCota:2011pn}. There
are two types of horizons in cosmology: the {\it causal} or particle horizon ($d_{H}$) and the {\it event} horizon
($d_{e}$). The former
determines the region of space which can be connected to some other region by causal physical
processes, at most through the propagation of light with $ds^{2}=0$. For the radiation cosmological era,
one has that $d_{H}(t)=2 t = H^{-1}$
and for the matter era one has $d_{H}(t)=3 t=2 H^{-1}$; $H^{-1}$ is sometimes called the Hubble horizon.
During inflation (under an exponential expansion of the Universe)
$d_{H}(t)= H^{-1} (e^{H t} -1)$ ($H=$const.) and hence, the causal horizon grows exponentially.
The event horizon, on the other hand, determines the region of space which will keep in causal contact
(again complying with $ds^{2}=0$) after some
time; that is, it delimits the region from which one can ever receive (up to some time $t_{\rm max}$)
information about events
taking place now (at time $t$). For the matter/radiation dominated eras
$d_{e} \rightarrow \infty$ as $t_{\rm max} \rightarrow \infty$. However, during inflation one has that
$d_{e} = H^{-1} (1 - e^{-(t_{\rm max}-t) H}) \approx H^{-1} $, which implies that any observer will see
only those events
that take place within a distance $\le H^{-1}$. In this respect, there is an analogy with black
holes, from whose surface no information can get away. Here, in an exponentially
expanding Universe, observers encounter themselves in a region
which is apparently surrounded by black holes \cite{GiHa77,Li90}, since they
receive no information located farther than $H^{-1}$.

Now, we turn back to the perturbation discussion. During the de Sitter stage the generation of perturbations, which is
a causal microphysical process, is localized in regions of the order of $d_{e} = H^{-1}$ in which
the microphysics operates coherently. At this time, the wavelength of inhomogeneities
grows exponentially (as the causal horizon does) and eventually they cross outside the event horizon. Much later on, they
re-enter into the event horizon, at the radiation and matter dominated epochs, to yield an almost scale invariant density
perturbation spectrum (Harrison-Zel'dovich, $n_{S} =1$), as is required for structure formation and measured
by different cosmological probes.

\begin{figure}[ht]
\includegraphics[width=12cm]{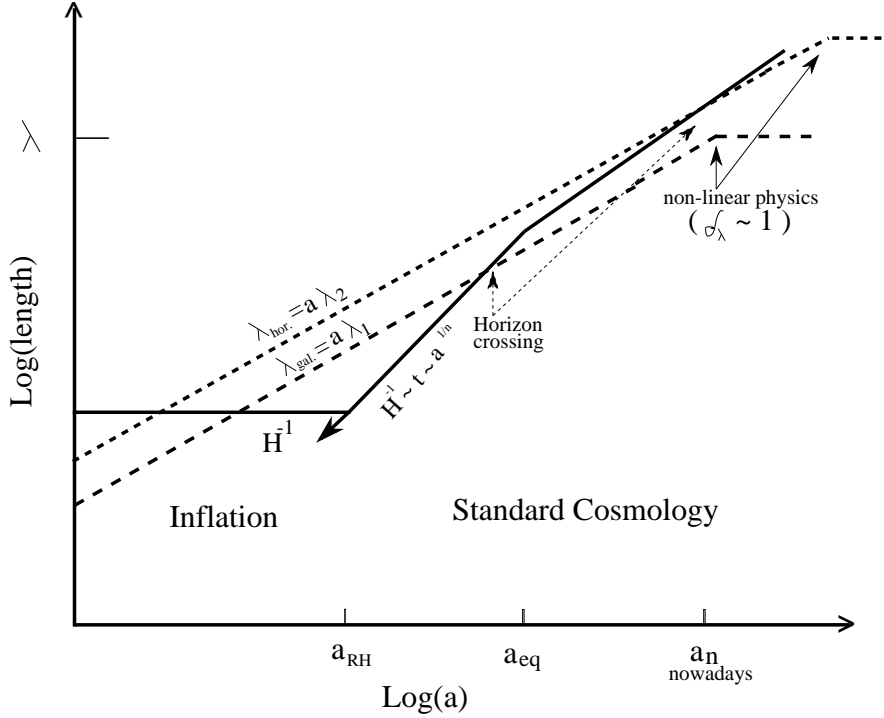}
\caption[Density perturbations in the inflationary cosmology]{Quantum
perturbations were initially subhorizon-sized. During
inflation they grow exponentially ($\lambda_{{\rm phys.}}=\lambda a(t)$),
whereas the event horizon remains almost constant. Then, eventually they
cross outside $H^{-1}$ and evolve as classical perturbations. Later on, they
re-enter the event horizon to produce an almost scale invariant,
Harrison-Zel'dovich density perturbation spectrum. In the figure are depicted two physical perturbations
scales: galaxy and horizon-sized. Figure adapted from Ref. \cite{KoTu90}.
\label{horin}}
\end{figure}

It was shown that the amplitude of inhomogeneities produced corresponds to the
Hawking temperature in the de Sitter space, $T_{H}=H/(2\pi)$.  In turn, this means that
perturbations with a fixed physical wavelength of size $H^{-1}$ are produced
throughout the inflationary era. Accordingly, a physical scale associated to
a quantum fluctuation, $\lambda_{{\rm phys}} = \lambda a(t)$, expands
exponentially and once it leaves the event horizon, it behaves as a metric
perturbation; its description is then classical, general relativistic. If
inflation lasts for enough time, the physical scale can grow as much as a
galaxy or horizon-sized perturbation. The field fluctuation expands always
with the scale factor and after inflation, it evolves according to $t^{n}$
($n=1/2$ radiation or $n=2/3$ matter). On the other hand, the Hubble horizon
evolves after inflation as $H^{-1}\sim t$. This means that it will come a time at
which field fluctuations cross inside the Hubble horizon and re-enters as
density fluctuations. Thus, inflation produces a gross spectrum of
perturbations, the largest scale ones being originated at the start of inflation with a
size $H^{-1}_{i}$, and the smallest ones with size $H^{-1}_{f}$ at the end of inflation (see Fig. \ref{horin}).

The power spectra for scalar ($S$) and tensor ($T$) perturbations are given by:
\begin{equation}
P_{S} (k) \approx  \left( \frac{H^{2}}{16 \pi^{3}\dot{\phi}_{c}^{2}}\right)
\;\vline{\atop{\atop {}_{k= a H}}},   \qquad
P_{T} (k) \approx  \left( \frac{H^{2}}{4 \pi^{2} m_{Pl}^{2} }\right)
\;\vline{\atop{\atop {}_{k= a H}}} ,
\label{dcp1in} \end{equation}
where $\dot{\phi}_{c}$ is the classical scalar field velocity. The equations are evaluated at the horizon
crossing ($k= a H$) during inflation. Each of the $k$-modes generate an anisotropy pattern in the CMBR
that was measured for scalar
perturbations by the COBE \cite{Sm92} and later probes.  The PLANCK satellite may have the chance
to detect the ratio of tensor to scalar amplitudes
$r\equiv C_{l}^{T}/ C_{l}^{S} < 0.12$ ($95\%$ limits)  \cite{Ade:2013zuv}, since the tensor modes
modulate CMBR photons coming from last scattering.

The power spectra above give rise to the observed curvature and tensor power spectra in terms of the
wavenumber ($k$) in a power law manner \cite{LyLi09}:
\begin{equation}
P_{\cal R} (k) = A_{S}  \left( \frac{k}{k_0} \right)^{n_{S} -1 +
\frac{1}{2} dn_{S}/d {\rm ln}k \: {\rm ln}(k/k_0) },   \qquad
P_{t} (k)   = A_{t}  \left( \frac{k}{k_0} \right)^{n_{t}}
\label{pp-n}
\end{equation}
that has been determined by recent CMBR probes, such as PLANCK to give a best fit of
$n_{S} = 0.96$ and $d n_{S}/d {\rm ln} k \approx -0.0090 $ \cite{Ade:2013zuv}. One should also have a tensor
spectral index $n_t$ that has not been measured yet.

These scalar and metric perturbations are small, but still very important. We discuss in the next section
how to include them
so that the information contained can be recognized and exploited.

\subsection{Perturbations inside the horizon}

We explained that in the early Universe baryons were tightly coupled to photons in an expanding background.
Baryonic and dark matter potential wells provoked the local collapse of density fluctuations up to a certain point,
at which the radiation pressure was big enough to pull out the matter apart and smooth the potential
wells. These oscillations of the plasma are in fact {\it acoustic waves}. As
we know, any wave can be decomposed into a
sum of modes with different wave numbers, $k = 2\pi/\lambda$.
Since  these modes are in the sky,  their wavelengths  are
measured as angles rather than as distances.
Accordingly, instead of decomposing the wave in a Fourier series, what is normally done is
to decompose the wave in terms of spherical harmonics, $Y_{lm}(\hat{n}) $, where $\hat{n}$ is the direction
of a measured photon.
The angular power spectrum can be expanded in Legendre polynomials,
since there is no preferred direction in the Universe and only the angular separation $\theta$ is relevant.
 A mode $l$  plays the same role of the wavenumber $k$, thus $l \approx 1/\theta$. We are
interested in the temperature fluctuations that are analyzed
experimentally in pairs of directions $\hat n$ and $\hat n'$, where $cos(\theta) = \hat n \cdot \hat n'$.
We then average
these fluctuations, obtaining the multipole expansions:
\begin{equation} \label{deltaT}
\frac{\Delta T}{T} = \sum_{l=1}^{\infty}\sum_{m=-l}^{l} a_{lm}(\vec{x},\eta)Y_{lm}(\hat{n}) ,
\qquad
P_{S}(\theta) = \sum \frac{(2 l +1)}{4 \pi} C_l P_l ({\rm cos}
\theta) ,
\end{equation}
where $P_{S}(\theta)$ is the angular power spectrum, $P_l$ are the Legendre polynomials, and the $C_l$
are estimated as averages of the $a_{l m}$ over $m$.
All this information can be used to determine the cosmological
parameters $\Omega_i$. We will not discuss here the
detailed calculations nor the curve that must be adjusted to obtain
the best fit values for such parameters.
The peak of the fundamental mode appears at approximately
\begin{equation} \label{ele}
l \simeq  \frac{200}{\sqrt{\Omega^{(0)}}} .
\end{equation}

BOOMERANG \cite{deB00} and MAXIMA \cite{Ha00} were
two balloon-borne experiments designed to measure the
anisotropies at scales smaller than the horizon at decoupling ($\theta_{\rm hor-dec} \sim 1  {}^{\circ}$),
hence measuring the
acoustic features of the CMBR. The sensitivity of the instruments allowed for a
measurement of the temperature fluctuations of the CMBR over a broad
range of angular scales. BOOMERANG found a value of $l = 197 \pm 6$ and MAXIMA-1 found a
value of $l \approx 220$. This implies that the cosmological density parameter $\Omega^{(0)} \approx 1$
[see Eq. (\ref{Omega})], suggesting
that the Universe is practically flat, $\Omega_{k}^{(0)} \approx 0$. These two experiments provided the
first strong evidence for a flat Universe from observations. Happily, this result was expected from
inflation since an accelerating dynamics effectively flattens the curvature of the event horizon, which
we later identify with our Universe (see
Fig. \ref{omein}). These
results were confirmed by WMAP in a series of data releases in the last decade, as well as by other
cosmological probes: the Universe is flat or pretty close to be flat. The problem in the exact determination
of the curvature is
because the CMBR anisotropies show strong degeneracies among the cosmological parameters
\cite{BoEfTe97,Zal97}.
However, the satellite PLANCK offers results on the density parameters with uncertainties less than
a percent level, $\Omega_{k}^{(0)} = -0.0105$  \cite{Ade:2013zuv}.

Since baryons and photons were in thermal equilibrium until recombination, also called
{\it last scattering} ($ls$), the
acoustic oscillations (BAO) were also imprinted in the matter perturbations, as they were in the CMBR
anisotropies. The sound horizon,
at the moment when the baryons decoupled from the photons, plays a crucial role in the determination
of the position of the baryon acoustic peaks. This time is known as the {\it drag epoch} which
happens at $z_{d} = 1/a_{d} -1$. The sound horizon at that time is defined in terms of the  effective
speed of sound of the baryon-photon plasma,
$c_{s}^{2}  \equiv \delta p_\gamma/ (\delta \rho_{\gamma} + \delta \rho_{b}) $,
\begin{equation}
r_{s} (z_{d}) = \int_{0}^{\eta_{d}} d \eta \, c_{s} (\eta)  = \frac{1}{3} \int_{0}^{a_{d}}
\frac{da}{a^{2} H(a) \sqrt{1+(3\Omega_{b}/4 \Omega_{\gamma})a}}  \, .
\end{equation}
Note that the {\it drag epoch} does not coincide with the last scattering. In most scenarios
$z_{ \rm d} < z_{ls}$ \cite{HuSu96}.
The redshift at the drag epoch can be computed with a fitting formula that is a
function of $\omega_{m}= \Omega_{m}^{(0)} h^{2}$ and $\omega_{b}= \Omega_{b}^{(0)} h^{2}$ \cite{EiHu98}.
The WMAP team, and recently PLANCK,
computed these quantities for the $\Lambda$CDM model, obtaining
 $z_{d} = 1059.29 \pm 0.65$ and  $r_{s} (z_{d}) = 147.53 \pm 0.64$ Mpc \cite{Ade:2013zuv}.

BAO can be characterized by the angular position and the redshift \cite{SeEi03,AmTs10}:
\begin{eqnarray}
\theta_{s} (z) &=& \frac{r_{s} (z_{d}) }{(1+z) \, d_{A}(z)}, \\
\delta z_{s} (z) &=&  r_{s} (z_{d}) \,  H(z),
\end{eqnarray}
where $d_{A}(z) = \frac{ 1}{H_{0}  \mid \Omega_{k} \mid^{1/2} (1+z)}
{\rm sin_k \left(\mid \Omega_{k} \mid^{1/2} \int^{z}_{0} \frac{dz'}{H(z')} \right)} $
is the proper (not comoving) angular diameter distance to the redshift $z$,
with ${\rm sin_k} = {\rm sin}$ for $\Omega_{k} <0$ and ${\rm sin_k} = {\rm sinh}$ for $\Omega_{k} >0$; where
$H(z)$ is determined by Eq. (\ref{hz}). The
angle $\theta_{s} (z) $ corresponds to the direction
orthogonal to the line-of-sight, whereas $\delta z_{s} (z)$ measures the fluctuations along the line-of-sight.
Observations
of these quantities are encouraging to determine both $d_{A}(z)$ and $H(z)$. However, from the
current BAO data is not simple to independently measure these quantities. This will certainly happen in
forthcoming surveys
\cite{Schelgel:2011zz}. Therefore, it is convenient to combine the two orthogonal
dimensions to the line-of-sight with the dimension along the line-of-sight to define \cite{Ei05}:
\begin{equation}
D_{V}(z) \equiv \left( (1+z)^{2} d_{A}(z)^2 \frac{z}{H(z)} \right)^{1/3}  \, ,
\end{equation}
where the quantity $D_{M} \equiv d_{A}/a = (1+z) d_{A}(z) $ is the comoving angular diameter distance.
The BAO signal has been measured in large samples of luminous red galaxies from the SDSS \cite{Ei05}.
There is a
clear evidence ($3.4 \sigma$) for the acoustic peak at a scale of $100 h^{-1}$ Mpc. Moreover, the scale and
amplitude of this peak are in good agreement with the prediction of the $\Lambda$CDM as confirmed by
the WMAP and PLANCK data. One finds that
$D_{V}(z=0.35)= 1370 \pm 64$ Mpc, and more recently new determinations of the BAO signal has been
published \cite{Carnero:2011pu} in which $\theta_{s}  (z=0.55)= 3.90{}^{\circ} \pm 0.38{}^{\circ} $
and $w= -1.03 \pm 0.16$ for the equation
of state parameter of the dark energy, or $\Omega_{M}^{(0)} = 0.26 \pm 0.04$ for the matter density, when the other
parameters are fixed. One also defines the BAO distance
$ d_{z} \equiv  r_{s} (z_{d}) /  D_{V}(z)$, which has been measured by surveys. For instance, an analysis of
the BOSS survey gives
$d(0.57) =13.67 \pm 0.22$ \cite{Anderson:2012sa}, which is the current most precise determination of the BAO scale.

Measuring the BAO feature in the matter distribution at different redshifts will help break the
degeneracy that exists
in the determination of the cosmological parameters. By combining line-of-sight with angular determinations
of the BAO feature
one will constrain even more the parameter space. Furthermore, a complete combination of BAO, the full matter
power spectrum, direct $H(z)$
measurements, supernovae Ia luminosities, and CMBR data will certainly help envisage the true nature of the
mysterious, dark Universe.

\section{Outlook}

We have reviewed the role that fluids have played in the entire history of the Universe. Their components are
relatively simple and behave
as perfect fluids, at least at the background level. The fluids' evolution is as follows: first, scalar fields
governed a very early inflationary dynamics with an equation of
state $w\sim -1$. After inflation, the Universe was deprived of particles and it had a very low temperature.
Then, reheating/preheating took place
to give rise to the hot Big Bang era, governed by a radiation period with $w = 1/3$. But the density of
radiation and/or relativistic particles (photons, neutrinos)
decayed faster than that of non-relativistic particles (protons, neutrons, DM) and eventually matter dominated
over the relativistic components in a dust ($w=0$) period
of the evolution. More recently, but still seven billion years ago, dark energy with $w\sim -1$ entered to
dominate the dynamics and to inflate the Universe again.

Real fluids are in a perturbed state, and the five main components of the Universe (photons, neutrinos, baryons,
dark matter, and dark energy) are not the exception. The plasma that composed the hot Big Bang era oscillated
with the well-known kinematics of perturbed fluids, and
as a consequence anisotropies in the CMB and inhomogeneities in the matter distribution left a unique
fingerprint that we measure at present. On
the other hand, if dark energy is the simplest candidate, the cosmological {\it constant}, its perturbations
are null, since it is simply a geometrical term in the Enstein's equations. But if it
is a fluid, perturbations are to be computed to understand their effect on structure formation.

Cosmological and astrophysical observations, since the early 1990's, have been playing a main role in
the cosmological science, which was governed mainly by
exact solutions and mathematical analyses. Indeed, we have just entered in a high precision era in which
the observations demand to construct new theoretical
observables, and vice versa. In the coming years, we expect not only to learn more about the fluids in
cosmology, such as dark matter and dark energy, but also
about the left-hand side of Einstein's equations: is GR correct? or, are modified gravity schemes more properly
fitted to the cosmic kinematics? These are quests
that challenge our present knowledge and that should be answered in the coming years.

\begin{acknowledgement}
This work was partially supported by the Consejo Nacional de Ciencia y Tecnolog\'{\i}a
of M\'exico (CONACyT) under the project CONACyT-EDOMEX-2011-C01-165873.
\end{acknowledgement}

\bibliographystyle{aipproc}   

\end{document}